\documentclass[
]{ceurart}

\usepackage[T1]{fontenc}
\usepackage{cleveref}
\usepackage{xcolor}
\usepackage{subcaption}
\usepackage{tabularx}
\usepackage{pifont}
\newcommand{\cmark}{\ding{51}}%
\newcommand{\xmark}{\ding{55}}%

\newcommand{\question}[1]{\textbf{#1}}
\newcommand{\answer}[1]{\textbf{#1}}

\begin{document}

\copyrightyear{2023}
\copyrightclause{Copyright for this paper by its authors.
  Use permitted under Creative Commons License Attribution 4.0
  International (CC BY 4.0).}

\conference{CHR 2023: Computational Humanities Research Conference, December 6
  -- 8, 2023, Paris, France}

\title{If the Sources Could Talk: Evaluating Large Language Models for Research Assistance in History}


\author[1]{Giselle \text{Gonzalez Garcia}}[%
orcid=0009-0008-2707-5420,
email=giselle.gonzalezgarcia@mail.concordia.ca,]
\cormark[1]
\fnmark[1]
\address[1]{Department of History and School of Irish Studies, Concordia University, Montreal, Canada}

\author[2]{Christian Weilbach}[%
orcid=0000-0001-7116-9338,
email=weilbach@cs.ubc.ca,
url=https://whilo.github.io/,
]
\fnmark[1]
\address[2]{Department of Computer Science, University of British Columbia, Vancouver, Canada}


\begin{abstract}
  The recent advent of powerful Large-Language Models (LLM) provides a new conversational form of inquiry into historical memory (or, training data, in this case). We show that by augmenting such LLMs with vector embeddings from highly specialized academic sources, a conversational methodology can be made accessible to historians and other researchers in the Humanities. Concretely, we evaluate and demonstrate how LLMs have the ability of assisting researchers while they examine a customized corpora of different types of documents, including, but not exclusive to: (1). primary sources, (2). secondary sources written by experts, and (3). the combination of these two. Compared to established search interfaces for digital catalogues, such as metadata and full-text search, we evaluate the richer conversational style of LLMs on the performance of two main types of tasks: (1). question-answering, and (2). extraction and organization of data. We demonstrate that LLMs semantic retrieval and reasoning abilities on problem-specific tasks can be applied to large textual archives that have not been part of the its training data. Therefore, LLMs can be augmented with sources relevant to specific research projects, and can be queried privately by researchers.
\end{abstract}

\begin{keywords}
  Artificial Intelligence (AI) \sep
  Machine Learning \sep
  Large Language Models (LLMs) \sep
  GPT \sep
  Historical Research Methods \sep
  Historical Writing
\end{keywords}

\maketitle

\section{Introduction \& Related Works}
\label{sec:introduction}

Researchers in History and the Humanities tend to accumulate thousands of papers, books, and other sources to be read, i.e. to be processed, in the near future~\citep{crane2006millionbooks}. However, only a handful of these end up in the bibliographical section of our papers, dissertations, and monographs. Because, we have to read them, i.e. process the information one page at the time. Yet, we still personally archive all these sources in our physical or digital libraries. Departing from this premise, we propose: what if these collections of academic texts can be incorporated into a corpus, infused into a Large Language Model (LLM), such as ChatGPT, and interrogated in new ways to produce better informed research outputs? This informed dialogue between researcher and machine would not necessarily be an output in itself, but it would significantly accelerate the research process in History and the Humanities. 

LLMs can be defined as a computer system that, given a word or text (token), can predict the words that would come after. Although these systems are not infallible, many researchers conceptualize the advent of LLMs as a proposition of full automation, especially in creative fields. In referring to full automation, we allude to the increasingly popular believe that LLMs entail complete replacement of humans in the production of knowledge and historical narratives. But considering the history of technology since the Industrial Revolution, machines have automated production processes that were often repetitive and exhausting for human labour. Automation freed human resources, it has provided humans with agency, and given us the possibility of dedicating ourselves to higher cognitive tasks~\citep{freyTechnologyTrapCapital2020}.

In the field of History, little attention has been devoted to understanding how using LLMs as part of our daily research praxis can have a deep impact and be a methodological game changer. LLMs have the potential to modify the way traditional historical archives are perused, the way primary sources are read and processed, the way theories and narratives are probed and validated, the way complex social processes are summarised, and finally, the way histories themselves are written. Our paper stems from the desire of incorporating LLMs into the authors' professional praxis in History and the Humanities. In doing so, we do not expect the LLMs to "tell the truth" for us, or to produce their own historical narrative in the form of synthetic new texts. Our main goal is to demonstrate what areas of the historical research process can be significantly enhanced and automated by having access to high-quality and accurate assistance. Even though, our professional experience is within the fields of History and Machine Learning, many of the case studies here exemplified can also appeal to other disciplines across the Humanities, and can illuminate the path forward in the field of Artificial Intelligence. 

When addressing the emergence of LLMs, several areas of concern are mentioned: (1). its potential abuse by students to plagiarize class assignments, (2). its incapacity to distinguish truth from falsehood \citep{kansteinerDIGITALDOPINGHISTORIANS2022}, (3). its environmental impact \citep{benderDangersStochasticParrots2021}, and (4) its ethical biases and risks \citep{kleinAreLargeLanguage2022}.  In the words of Wulf Kansteiner, "Large language models such as GPT-3 are able to generate compelling, non-plagiarized texts in response to simple natural language inputs, thus providing students with an opportunity to produce high-quality written assignments with minimum effort." \citep{kansteinerDIGITALDOPINGHISTORIANS2022}. One of the main limitations of LLMs (specifically of GPT-3), that Kansteiner identified, is its structurally inability "to attribute the statements it generates to a specific textual origin, let alone assess the factual reliability of any of its textual inputs or outputs."~\citep{kansteinerDIGITALDOPINGHISTORIANS2022}. This is a problem to which our paper offers a tentative solution by arguing that combining an LLM with carefully selected corpora made of primary and secondary sources reduces the margins for errors, and produces texts that have an origin in peer-reviewed sources. A similar approach has been already implemented by \citet{manjavacasarevaloMacBERThDevelopmentEvaluation2021} to positive results in the form of the MacBERTh model. But, unlike them, we do not train a new model, but augment pre-trained ones.

Moreover, the answers we intend on retrieving from the LLMs include specific details about its sources in the form of text chunks. Therefore, instead of getting answers to our questions from models pre-trained on data only acquired from the world wide web, we propose to prompt the models with a context of chunks that have already been through the established process of academic validation, which can also be checked to validate the response. This approach also shows promise to train better LLMs in general~\citep{gunasekarTextbooksAreAll2023}.

\citet{benderDangersStochasticParrots2021}, present a critical overview of the most problematic aspects of LLMs, including their significant contributions to climate change, and their tendency to reproduce and validate hegemonic ideas that are over-represented on the internet, and therefore on the training data that infuses these models. At the same time LLMs show promise to scale deliberation and integrate diverse view points in democratic processes~\cite{smallOpportunitiesRisksLLMs2023}. We do not intend to add to this debate at this early stage in our research, but we acknowledge the current academic debate that problematizes the multiple societal paradoxes that the advent of LLMs (ChatGPT, in particular) have brought to the forefront.    

\begin{figure}[t]
    \centering
    \includegraphics[width=0.75\textwidth]{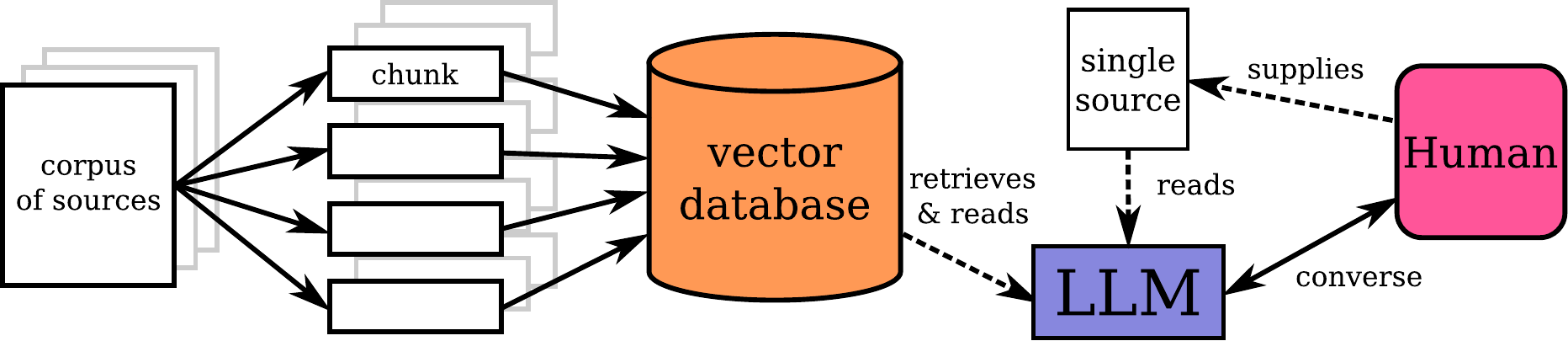}
    \caption{Interaction between a Human and an LLM in form of a conversation. The human optionally supplies a single source text document for the LLM to read first. The LLM can also retrieve text chunks from a vector database filled from a corpus of source documents. Dashed lines denote optional dependencies.}
    \label{fig:overview}
\end{figure}

To evaluate such an scenario of historical research assistance empirically, we implemented an open-source pipeline (\Cref{fig:overview}) with different freely available LLMs that can be run locally (and privately) by any researcher (\Cref{sec:methodology}). We call the implementation KleioGPT. We then compiled a comprehensive academic corpus of digitized history monographs from the first author's academic research in Irish Migration Studies, and conducted a case study from the perspective of a professional historian in~\Cref{sec:evaluation}. Our case studies consist of two important textual research tasks, (1). question-answering in~\Cref{sec:academic_question_answering} and (2). data extraction in \Cref{sec:extracting_and_organizing_data}. We assembled 40 prompts for the question-answering task and graded the answers given by different LLMs. We took into account whether the LLMs really responded to our prompt correctly, and paid attention to which sources from our corpus were pulled for answering correctly. For testing data extraction we resorted to the 9 volume genealogical collection \emph{Historia de Familias Cubanas} by \citet{santacruzymallenHistoriaFamiliasCubanas1940b}.

Since the texts from the different corpora are fed to the LLMs in their general textual representation, any other form of document retrieval from an archive, such as full text, web or metadata search would equally work. We have not explored this in this work.

\section{Background}
\label{sec:background}

Our pipeline shown in~\Cref{fig:overview} consists of a two stage process. In the first stage, the textual content of all documents in the archive is extracted and sliced into equally-sized overlapping chunks. These chunks are then individually stored into a vector database from which they can be efficiently retrieved later. To answer a question, first a number of relevant chunks are queried from the vector database, and then optionally fed to the LLM together with the question as its prompt. This allows the LLM to answer archive-specific questions that exceed the information of its training data.

\subsection{Vector databases} 
\label{sec:vector_databases}

Vector databases index so called ``embeddings''. These embeddings summarize the content of a text chunk from a document in a fixed length vector of real numbers. To create the database, first each chunk is paired with its respective embedding. The chunk is then added under its embedding to the index of the database. Once embedded, these vectors have the property of locating semantically similar documents near each other. While executing a query, the database refers to its index to find those documents located in proximity to the embedding that matches the text of the query. It then retrieves the text chunk of each nearby embedding from the archive.  

\subsection{Large language models}
\label{sec:large_language_models}

\begin{table}[t]
    \centering
    \caption{The different LLMs used in our experiments. Settings annotated with ? are not documented publicly. GPT3 was originally described as having 175 billion parameters \cite{brownLanguageModelsAre2020}. The ChatGPT and GPT3 results were created in July 2023.}
    \begin{tabular}{l|c|c|c|c|c}
         \textbf{model}      & \textbf{private} & \textbf{\#parameters} & \textbf{context size} & \textbf{access} & \textbf{Hugging Face id}\\
         \hline
         Falcon \cite{falcon40b} & \cmark & 7b &    4k & langchain & \texttt{falcon-7b-instruct} \\
         \hline
         XGen \cite{XGen}   & \cmark & 7b &    8k & langchain & \texttt{xgen-7b-8k-inst} \\
         \hline
         Beluga \cite{mukherjeeOrcaProgressiveLearning2023, touvronLlamaOpenFoundation2023}    & \cmark & 7b &    4k & langchain & \texttt{StableBeluga-7b} \\  
         \hline
         GPT3 \cite{brownLanguageModelsAre2020} & \xmark & 175b? &   4k & langchain & \texttt{gpt-3.5-turbo} \\
         \hline
        ChatGPT \cite{brownLanguageModelsAre2020}  & \xmark & 175b? & 4k? & web & - \\
    \end{tabular}
    \label{tab:llm_table}
\end{table}

Large language models are a form of generative model. Generative models are probabilistic models that can be fit to a data set and  generate synthetic data that mimics it.  In our work, we depart from the successful class of generative pre-trained transformer (GPT) models~\cite{brownLanguageModelsAre2020, vaswaniAttentionAllYou2017a}. GPT models are pre-trained on large public and private archives to predict the next word in a given context. This type of pre-training has proven to be effective in yielding state-of-the-art language models that perform well when given tasks of interest for researchers in the Humanities~\citep{hosseiniNeuralLanguageModels2021,manjavacasarevaloMacBERThDevelopmentEvaluation2021}. These include question-answering and summarization, a comprehensive recent survey is provided in \citet{yangHarnessingPowerLLMs2023}. 

While pre-training such a model on billions of words is very costly, both in terms of time and compute, running these pre-trained models is possible on commodity hardware. For our work, we selected competitive (at the time of their release) publicly downloadable models, as well as the most popular models, such as GPT3 and ChatGPT; see \Cref{tab:llm_table}. For the open models we chose small versions that can be run on consumer graphics cards (\Cref{app:experimental_details}).
The primary interface to LLMs is the so called prompt, which is the text provided to the model to describe the context and task at hand. The model then appends its answer to the prompt as a form of auto-completion by next word prediction. We document our prompts for each experiment in \Cref{app:question_answer_results}. Generation in LLMs is generally stochastic and the amount of stochasticity can be modulated by a so called temperature. A common problem in LLMs is so called hallucination (see \Cref{app:hallucination_example}), which refers to the fact that they can make up answers that have no factual basis. 

\section{Methodology}
\label{sec:methodology}

We integrated available LLM open-source software to execute the task described in \Cref{sec:retrieval_augmented_question_answering} together with an interactive question-answering mode in a Python code base for KleioGPT.\footnote{\url{https://github.com/GissyGonzalez/KleioGPT}} For \Cref{sec:extracting_and_organizing_data} we directly prompt ChatGPT as described there.

\subsection{Retrieval augmented question-answering}
\label{sec:retrieval_augmented_question_answering}

Our methodology is based on the established combination of memory retrieval from vector databases with LLMs~\cite{borgeaudImprovingLanguageModels2022}. Augmenting sources does not only help the LLM to answer the question, it also provides a way to check whether the answer can be backed up by the documents in the vector database, a step we think is necessary in an academic setting. Our implementation uses a pre-trained transformer embedding model,\footnote{\url{https://huggingface.co/sentence-transformers/all-MiniLM-L6-v2}} the vector database Chroma,\footnote{\url{https://docs.trychroma.com/}} and, to integrate the different LLMs, we use LangChain.\footnote{\url{https://langchain.com/}} For a fair and reproducible comparison we set the temperature of all models (except for ChatGPT where we have no control) close to zero ($10^{-5}$), rendering generation effectively deterministic. A zero temperature corresponds to picking the most likely answer in each step and encourages factual correctness over creativity. We chose the default settings of each model in LangChain for all other parameters. We improved the retrieval mechanism of LangChain to filter out chunks of text that are from bibliographic sections of the sources and chunks that contain less than 200 characters.
During question-answering, we loaded a set of questions from a prepared comma separated value (CSV) file. In return, we obtained another CSV file as a report with the initial questions, the answers, and the retrieved sources. Examples are shown in \Cref{sec:academic_question_answering}. This simple process both lends itself to automated batch-processing of questions and is accessible to non-technicians through its tabularized inputs and outputs. 

\subsection{Datasets}
\label{sec:dataset}

\subsubsection{Migration Studies Corpus}
\label{sec:bibliographical_dataset}

To examine the question-answering capacity of each LLM, we assembled a corpus that forms our Migration Studies dataset. Our aim is to show that researchers can assemble their own corpora tailored to their specific projects. This bibliographic corpus is made of 86 books from leading scholars in Irish Migration Studies, Cuban Studies, general Migration Theory, histories of Slavery, etc. They have been selected from the main fields of study of the first author, and are the basis for their doctoral research. This corpus is representative of the best works in these fields, and among its authors are the leading academic experts in each subject.

In order to verify (and grade) the answers provided by each LLM, we selected volumes the first author was familiar with, had read recently, and had a knowledge of each writer's placement within the historiography and the broader academic conversation in themes that include migration, race, gender, and slavery. These are mostly historical monographs that have been through a peer-review publication process. This does not exempt them from reproducing the societal biases of their authors, but we believe no text is exempt of reproducing the ideologies and cultural systems researchers subscribe to. Nonetheless, we prioritized feeding KleioGPT sources that have gone through a well-established academic validation process, and can be generally factually trustworthy. We provide a list of these sources in \Cref{app:irish_migration_studies_dataset}. 

As mentioned above, researchers can gather their own corpus and ingest them into their own versions of KleioGPT. Every text loaded into our experiments was in PDF format. We made sure that each PDF had a plain text layer that ensured its content is machine readable. Users should make sure that their OCR'ed texts are as clean as possible for better results. Our Migration Studies corpus is exclusively in the English language, although this methodology can be applied  to corpora in other languages if the embeddings and LLMs support it.

\subsubsection{Family History Books Compilation}
\label{sec:family_history_books_compilation}

\begin{figure}[t]
    \centering
    \begin{subfigure}[b]{0.45\textwidth}
    \includegraphics[width=\textwidth]{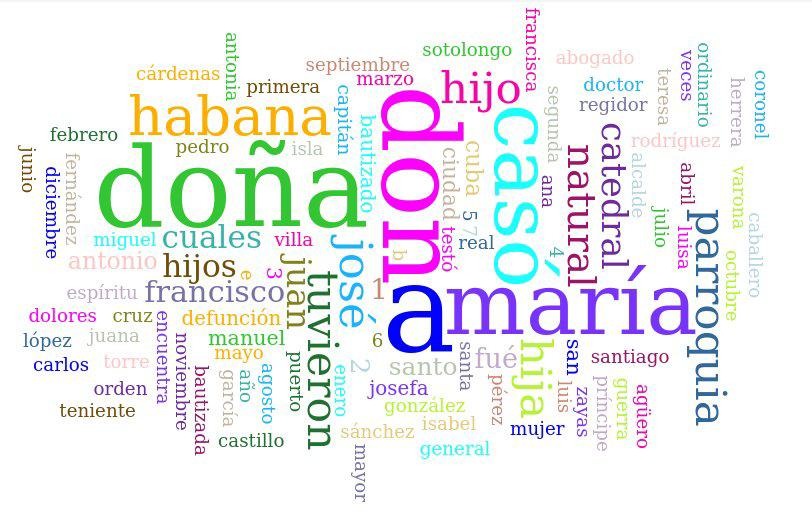}
    \caption{Word cloud of the 9 volumes of \emph{Historia de Familias Cubanas} by Francisco Javier de Santa Cruz \citep{santacruzymallenHistoriaFamiliasCubanas1940b} showing the word frequency and the regularity of names and expressions. Made by the authors with VoyantTools.}
    \label{fig:wordcloud}        
    \end{subfigure}
    ~
    \begin{subfigure}[b]{0.45\textwidth}
    \includegraphics[width=\textwidth]{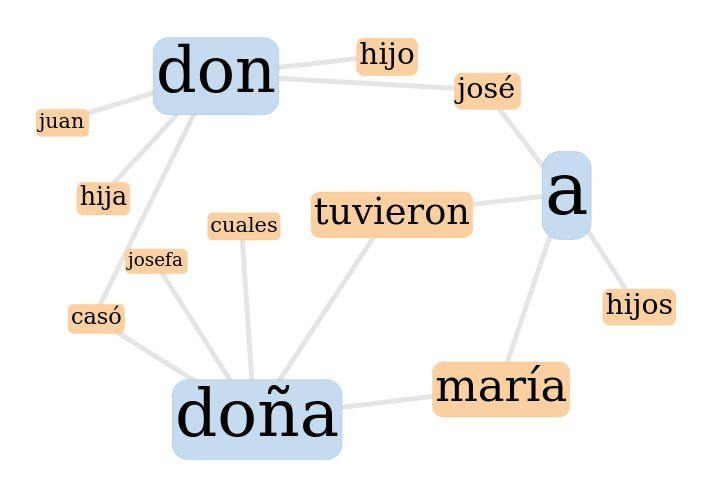}
    \caption{Links visualization of the 9 volumes of \emph{Historia de Familias Cubanas} by Francisco Javier de Santa Cruz showing the regular structure of genealogical language corpus. Made by the authors with VoyantTools.}
    \label{fig:genealogical_language}        
    \end{subfigure}
    \label{fig:family_history_book_compilation}
\end{figure}

In the past five years a considerable part of the first author's doctoral research on Irish migrants in the Hispanic Caribbean has consisted of reviewing genealogical compilations in search for biographical data related to 19th-century migrants. In many occasions, genealogical data is the only type of information retrievable to rescue the histories of this population. 

Genealogical sources are usually structured in a regular pattern that outlines birth, marriage, death facts, and kinship connections among people. These are worded similarly. Automating the extraction of this type of data from family history book compilations is a task that many historians and researchers in the Humanities undertake frequently. It is labour intensive, repetitive, and consumes significant amounts of time that could be devoted to other research endeavours. Therefore, the possibility of automation by using LLMs is a very attractive one. 

Our genealogical dataset is made of the 9 volumes of \emph{Historia de Familias Cubanas} \citet{santacruzymallenHistoriaFamiliasCubanas1940b} (including OCR). This corpus consists of 1,803,596 total words and 39,114 unique word forms. The data set is visualized in \Cref{fig:family_history_book_compilation}.

\section{Evaluation}
\label{sec:evaluation}

\subsection{Academic question-answering}
\label{sec:academic_question_answering}

To interrogate our corpus from \Cref{sec:academic_question_answering} we designed a fixed set of questions and followed the methodology outlined in \Cref{sec:retrieval_augmented_question_answering}. Our goal was to examine and compare the quality of different LLMs responses to four different types of queries: factual, argumentative, descriptive, and integrative. Their level of complexity corresponded to the types of questions undergraduate student in the 200 and 300 level courses would typically face. We graded the answers in a pass/failed binary, giving one point for correct answers, and zero for the opposite. We also knew beforehand the texts contained accurate answers to these questions. 
All the questions can be found in \Cref{app:question_answer_results}.

\begin{table}[t]
    \centering
    \caption{Results of the question-answering task. The percentages are averaged over the 10 questions in each category. Percentages in bold are the best performing in each category.}
    \begin{tabular}{l|c||c|c|c|c|c}
         \textbf{model}     & \textbf{\#chunks} & \textbf{factual} & \textbf{argumentative} & \textbf{descriptive} & \textbf{integrative}  & \textbf{total} \\
         \hline
         Falcon & 0 &    60\% &          60\% &        40\% &        70\% & 57.5\% \\
                            & 4 &    30\% &          60\% &        40\% &        70\% & 50\% \\
                            & 8 &    20\% &          70\% &        80\% &        70\% & 60\% \\
         \hline
         XGen    & 0 &    80\% &         70\%  &     70\% &       80\% & 75\% \\
                            & 4 &    90\% &         \textbf{100\%} &       80\% &  80\%  & 87.5\% \\
                            & 8 &    80\% &         90\%  &        90\% & 60\% & 80\% \\
        \hline
        Beluga     & 0 &    80\% &         80\%  &     90\% &       60\% & 77.5\% \\
                            & 4 &    80\% &         50\% &      90\% &       90\%  & 77.5\% \\
                            & 8 &    90\% &         70\%  &     \textbf{100\%} &      \textbf{90\%} & 87.5\% \\        
        \hline
         GPT3       & 0 &   70\% &        70\% &  \textbf{100\%} &   60\% & 75\% \\
                            & 4 & \textbf{100\%} & 60\% &       70\% &       70\%  & 75\% \\
                            & 8 &   90\% &         60\% &  \textbf{100\%} & 80\% & 82.5\% \\
         \hline
        ChatGPT             & 0 & \textbf{100\%} &         80\%  & \textbf{100\%} & \textbf{90\%} & \textbf{92.5\%}\\
    \end{tabular}
    \label{tab:qa_result_table}
\end{table}

We summarize our results in \Cref{tab:qa_result_table}. All LLMs could answer the majority of questions in all configurations. The retrieved text chunks from the vector database in general improve the results and improve important context to validate the answer, but the LLMs are interestingly already able to answer many questions even without any sources. ChatGPT performed best even without any sources and remarkably we found that there is a big gap to GPT3 accessed through the OpenAI API, which is underlying ChatGPT. There are probably significant extensions applied to the web interface of ChatGPT. Both XGen and Beluga consistently outperform GPT3 and both are competitive with ChatGPT.

Acting under 0 chunks, i.e. without having access to our corpus, the LLMs tended to fail when giving answers to specific questions like: "What was the population of Ireland in 1841?". They were also not able to summarize the contents and main ideas of books like Kerby A. Miller's \emph{Emigrants and Exiles} \citep{millerEmigrantsExilesIreland1985}.
Without the sources in our dataset, LLMs like Falcon were still able to define conceptual ideas and answer correctly prompts like:  "Can you define what is migration/Irish nationalism?". Overall, tasks that required the LLMs to generate its own ideas or argumentation (like those in the integrative section of our Q\&A) tended to perform worse, or even prompted the LLMs to hallucinate (see \Cref{app:hallucination_example}) completely off-the-mark answers, when it did not have the backup of the corpus. 

\paragraph{Factual} The ten questions dedicated to retrieval of factual information focused on obtaining precise data such as dates, quantities, locations, etc. Fact-checking queries are very typical during the historical research process. When these types of question are posed to search engines like: Google, Bing, Startpage, etc., these point us to a list of often millions of indexed results. We have to open web pages to retrieve the information we are looking for and evaluate it. As a system to effectively retrieve factual data, LLMs give us a direct answer, therefore shortening significantly the time invested on this task. These types of questions were easy to grade because they were either right or wrong. When some answers were an approximate, results were graded according to the standard academic consensus. 

\begin{quote}

\question{Question 5}: "Where in Ireland are Protestants found?"

\answer{XGen}: "Protestants were found in all parts of Ireland, but they were particularly concentrated in urban areas and in the southern province of Ireland. In the early 1800s, they were a majority in County Londonderry."

\answer{Falcon}: "Protestants are found in all parts of Ireland, but are more commonly found in urban areas."

\answer{GPT3}: "Protestants are found throughout Ireland, but they tend to be concentrated in towns and urban areas."

\answer{ChatGPT}: "Protestants in Ireland can be found throughout the country, but their distribution varies across different regions. Historically, Protestants have had a stronger presence in Northern Ireland, particularly in the counties of Antrim, Down, Armagh, and Londonderry (also known as Derry). In these areas, Protestant communities, including various denominations such as Anglicans, Presbyterians, and Methodists, have traditionally been more prominent. However, it's important to note that there are Protestant communities and individuals living in all parts of Ireland, including the Republic of Ireland."

\end{quote}

All models answer \question{Question 5} correctly, but ChatGPT provides a lot more relevant details. It is unclear why it behaves so differently, but it could be that it is provided an additional prompt that is not publicly accessible.

\paragraph{Argumentative} We also designed ten questions for this type of query. In this category, we asked for summaries of concepts, ideas, and entire books. It is important to clarify that, once our corpus was added, LLMs summarize books based on other scholars opinion's of those books retrieved from the chunks, and not on their own complete reading of each book.\footnote{The recent release of Claude 2 (\url{https://claude.ai}) with a large context window of 100.000 tokens provides such functionality, unfortunately it was not yet programmatically available at the time of submission to the authors.}  Essentially, it will summarize what other authors said about a specific book, and not the book itself. This can be valuable to realize what the academic consensus is about a specific work, idea, or theory; and to identify contentious arguments about them. Most questions in this category that the LLMs failed, were because they hallucinated or fabricated events that were easily identifiable as factually wrong.

\begin{quote}

\question{Question 3}: "Can you summarize the three main arguments in Kerby Miller’s book Emigrants and Exiles?"

\answer{XGen}: "1. Irish American nationalism is the primary expression of a powerful sense of exile among Irish immigrants to America. 2. Irish immigrants to America were deeply homesick and resistant to the culture of individualism. 3. Irish chain migration to North America was not passive, but rather relied on the recommendations of previous migrants." 

\answer{GPT3}: "I don't know the answer." 

\answer{ChatGPT}: "Kerby Miller's book "Emigrants and Exiles: Ireland and the Irish Exodus to North America" explores the causes, experiences, and consequences of Irish emigration to North America during the 18th and 19th centuries. While it is challenging to fully summarize all the arguments in the book, here are three key themes..." 

\end{quote}

In the example above, the authors were expecting a succinct answer similar to the one provided by XGen. GPT3's reply was unexpected given that Miller's book is a classic volume in the field. \emph{Emigrants and Exiles} (1985) has been discussed in length on the web, and according to Google Scholar, is cited 1663 times. Discussions about this book should be on GPT3 training data. Its response contrasts starkly with ChatGPT's lengthy and detailed one. 

\paragraph{Descriptive} Raising the level of complexity, in this category, we included questions that necessitated more elaborated answers about historical events and processes. 

\begin{quote}

\question{Question 4}: "Can you describe the events that led to the Great Irish potato Famine?"

\answer{ChatGPT}: "The Great Irish Potato Famine, also known as the Great Hunger or the Irish Potato Famine, was a devastating period in Irish history that occurred between 1845 and 1852. It was a catastrophic event primarily caused by a combination of potato crop failure, economic policies, and socio-political factors. Here is a description of the events that led to the famine: Potato Dependence: ..." 

\end{quote}

All models answered this question correctly. This can be attributed to how pervasive in society the topic at hand is and how much it has been discussed online. XGen again gave a correct concise answer, however we were expecting a lengthier description given the complexity and gravity of the Great Famine. ChatGPT not only gave a very detailed answer, but it also outlined the causal chain of events, step by step, leading to the famine and its extent (the full answer is in \Cref{app:chatgpt_answer_famine}). Unlike ChatGPT, the other LLMs provided answers that were below the standard of what would have been expected of an early undergraduate Irish Studies student.  

\paragraph{Integrative} To elaborate this type of questions, we took inspiration in the type of exercises graduate students in North America are tested on during their comprehensive examinations. In answering these questions, the LLMs needed to demonstrate more creativity than in the previous ones. This section was supposed to entail the highest level of difficulty to be tested during our experiments. Here, the LLMs struggled with replying in accordance to the level expected of a human graduate student in History. 

\begin{quote}

\question{Question 3}: "What is the future of History as a discipline?" 

\answer{XGen}: "The future of history as a discipline is uncertain and likely to be shaped by a variety of factors, including changes in technology, the rise of new historical methodologies, and shifts in the broader cultural and political landscape. Some experts predict that history will continue to evolve and become more interdisciplinary, incorporating insights from other fields such as anthropology, sociology, and economics. Others suggest that the rise of digital humanities and the increasing availability of large datasets may lead to new approaches to historical research and analysis. Additionally, the future of history may be influenced by broader trends such as the increasing importance of global and transnational history, as well as the growing interest in the history of science and technology. Ultimately, the future of history will depend on the choices and priorities of historians and the broader academic community." 

\answer{GPT3}: "Based on the given context, it is not possible to accurately determine the future of history as a discipline. The context provides information about the challenges and changes that the discipline of history has faced, such as the questioning of the nature of history, the interaction with humanities, arts, and sciences, and the interdisciplinary nature of new scholarship. However, there is no explicit information or discussion about the future trajectory of the discipline."
   
\end{quote}

In the example provided, we observe the contrast between two different approaches: one in which GPT3 determined it had not been provided enough material from the sources and is therefore uncertain about how to answer the question; and another in which XGen based on the same materials, tried to give a well-rounded reply. 

\subsection{Extracting Tabular Data from Genealogical Sources}
\label{sec:extracting_and_organizing_data}

\begin{figure}[ht]
    \centering
    \includegraphics[width=0.7\textwidth]{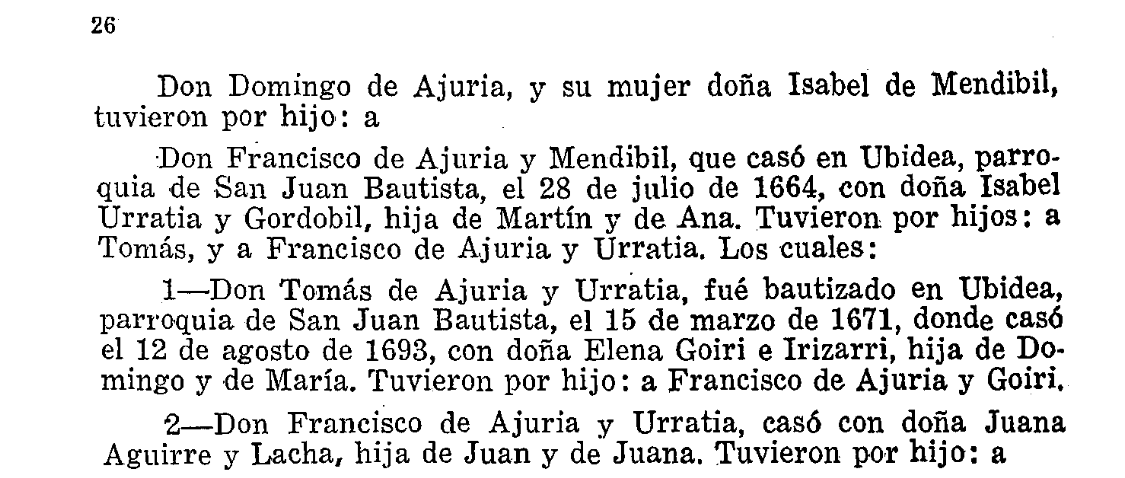}
    \caption{An screenshot of a page from \emph{Historia de Familias Cubanas} describing the descendants of Domingo de Ajuria (\Cref{sec:family_history_books_compilation}).}
    \label{fig:name_excerpt_example}
\end{figure}

\begin{table}[ht]
\caption{The above table (broken in three lines) is extracted from \Cref{fig:name_excerpt_example} automatically by ChatGPT. The date highlighted in red was wrongly extracted from baptism date. The text fragments in \textbf{bold} were in the text, but missing in the table and manually added.}
\centering
\begin{tabular}{|l|l|l|l|l|}
\hline
\textbf{Full Name} & \textbf{Date of Birth} & \textbf{Place of Birth} & \textbf{Baptism Date} & \textbf{Marriage Date} \\ \hline
Domingo de Ajuría & Unknown & Ubidea & Unknown & Unknown \\ \hline
Isabel de Mendibíl & Unknown & Unknown & Unknown & Unknown \\ \hline
Francisco de Ajuria \textbf{y Mendibil} & Unknown & Ubidea & Unknown & \textbf{28-07-1664} \\ \hline
Isabel Urratia \textbf{y Gordobil} & Unknown & Unknown & Unknown & \textbf{28-07-1664} \\ \hline
Tomás de Ajuria \textbf{y Urratia} & \textcolor{red}{15-03-1671} & Ubidea & \textbf{15-03-1671} & 12-08-1693 \\ \hline
Elena Goiri e Irizarri & Unknown & Unknown & Unknown & 12-08-1693 \\ \hline
Francisco de Ajuria \textbf{y Goiri} & Unknown & Unknown & Unknown & Unknown \\ \hline
\end{tabular}\\
\vspace{0.2cm}
\begin{tabular}{|l|l|l|}
\hline
\textbf{Father's Full Name} & \textbf{Mother's Full Name} & \textbf{Children's Full Name} \\ \hline
Unknown & Unknown & \textbf{Francisco de Ajuria y Mendibil} \\ \hline
Unknown & Unknown & Unknown \\ \hline
Domingo de Ajuría & Isabel de Mendibíl & Tomás de Ajuria; \textbf{Francisco de Ajuria} \\ \hline
\textbf{Martin} & \textbf{Ana} & Tomás de Ajuria; \textbf{Francisco de Ajuria} \\ \hline
Francisco de Ajuria & Isabel Urratia & Francisco de Ajuria \textbf{y Goiri} \\ \hline
\textbf{Domingo} & \textbf{Maria} & Francisco de Ajuria \textbf{y Goiri} \\ \hline
Tomás de Ajuria & Elena Goiri e Irizarri & Unknown \\ \hline
\end{tabular}\\
\vspace{0.2cm}
\begin{tabular}{|l|l|}
\hline
\textbf{Spouse's Full Name} & \textbf{Gender} \\ \hline
\textbf{Isabel de Mendibil} & Male \\ \hline
Domingo de Ajuría & Female \\ \hline
\textbf{Isabel Urratia y Gordobil} & Male \\ \hline
Francisco de Ajuria & Female \\ \hline
Elena Goiri e Irizarri & Male \\ \hline
Tomás de Ajuria \textbf{y Urratia} & Female \\ \hline
Unknown & Male \\ \hline
\end{tabular}
\label{tab:table_extraction}
\end{table}

In this section, we took several pages from \emph{Historia de Familias Cubanas} (\Cref{sec:family_history_books_compilation}) and manually provided their full contents to the web interface of ChatGPT (the original text can be seen in \Cref{fig:name_excerpt_example}). The content of these 9 volumes is entirely in Spanish. Even though only 7\% of GPT3 training data is in a language other than English~\citep{brownLanguageModelsAre2020}, ChatGPT performed well in Spanish, translating accurately the excerpts provided. Afterwards, we designed the following prompt: 

\begin{quote}
    \textbf{Prompt example}: "From the previous text, list all the names of people in a table with columns: full name of each person, relationship, date of birth (format DD-MM-YYYY), place of birth, date of death (format DD-MM-YYYY), baptism date  (format DD-MM-YYYY), marriage date  (format DD-MM-YYYY), place of residence, full name and surname of father, full name and surname of mother, full name and surname of children, full name and surname of spouse, and occupation. Try to infer the gender of each person, and add a column Gender."
\end{quote}

In \Cref{tab:table_extraction} we show that ChatGPT is able to extract data from \emph{Historia de Familias Cubanas} (9 volumes) in \citet{santacruzymallenHistoriaFamiliasCubanas1940b} into a structured tabular format by simply prompting it to create a table with the respective column names. 

A large part of the information is correctly mapped into the table. However, ChatGPT did miss some of the information highlighted in bold. The missing fragments show regularities such as family names that are connected by the conjunction ``y''. We believe that it is likely that by instructing ChatGPT explicitly on these expressions, e.g. by giving example translations or erroneous lines, the LLM would miss less data. We plan to address this in future work together with an automatic page wise extraction of the tables with our prompt.

As can be seen in \Cref{fig:name_excerpt_example}, the gender information was not in the original text. We asked ChatGPT to infer it automatically based on its pre-trained knowledge about Spanish name conventions.

Out of all the dates extracted, there is only one birth date (in red) inferred wrongly, everything else is correct. If this low error rate reflects the performance on the full dataset, then a large amount of data could be extracted into a tabular representation just by asking ChatGPT to do so. We could not get the same extraction quality out of the other LLMs, but expect them to catch up in the near future.

\section{Conclusion}
\label{sec:conclusion}

In this paper we have demonstrated that LLMs can provide a fluent conversational research assistance while being sufficiently accurate for an academic environment. While open source models that fit onto commodity hardware are not yet competitive with ChatGPT in all tasks, they already provide a private alternative on question-answering tasks on sensitive data. While this work already highlights the potential of LLMs for historical research, problems such as hallucination and biases in the training data will also probably be significantly reduced in the near future. Since the field of LLMs is evolving very quickly, we think that very soon much better open source LLMs will be available. We will continue to design tests to evaluate the LLMs abilities in the near future.

Interesting avenues for future work are better integration of memory to both help the generation of the LLMs and make it easier to validate them against the sources retrieved from memory. Models such as Claude 2 make use of larger context windows for better extraction and summarization abilities and show a lot of promise in our ongoing evaluations as well. We think that open source solutions that facilitate research will be widely available to researchers in the Humanities including models that can be run on machines without GPUs.\footnote{An older example for this is \url{https://gpt4all.io/index.html}}

\section*{Acknowledgement}
\label{sec:acknowledgment}

We acknowledge the support of the Natural Sciences and Engineering Research Council of Canada (NSERC), the Canada CIFAR AI Chairs Program. This research was enabled in part by technical support and computational resources provided by the Digital Research Alliance of Canada Compute Canada (alliancecan.ca), the Advanced Research Computing at the University of British Columbia (arc.ubc.ca), Amazon, and Oracle.

\bibliography{bibliography}

\newpage

\appendix

\section{Experimental details}
\label{app:experimental_details}

We ran experiments for the local LLMs StableBeluga-7b, falcon-7b-instruct and xgen-7b-8k-inst on an A5000 NVIDIA GPU. 

\section{Academic question-answering}
\label{app:question_answer_results}

\subsection{Questions}

\subsubsection{Factual}

\begin{enumerate}
    \item What was the population  of Ireland in 1841?
    \item When did Irish Potato Famine start?
    \item When did Irish Potato Famine finish?
    \item When did the Irish Parliament past the Act of Union?
    \item Where in Ireland are Protestants found?
    \item Where in Ireland are native Irish speakers found?
    \item What was the population  of Ireland in 1861?
    \item When did Catholic Emancipation happen?
    \item What happened in Ireland in 1848?
    \item When was the Land League created?
\end{enumerate}

\subsubsection{Argumentative}

\begin{enumerate}
    \item Can you make a definition of migration in ten sentences?
    \item What is Irish nationalism?
    \item Can you summarize the three main arguments in Kerby Miller’s book Emigrants and Exiles?
    \item Can you summarize the three main arguments in Donald Akenson’s book The Irish in Ontario?
    \item What is the Irish diaspora?
    \item What is Irish unionism?
    \item What are three differences between Irish nationalism and Irish unionism?
    \item What are three similarities between Irish nationalism and Irish unionism?
    \item Can you summarize what David Fitzpatrick’s book Oceans of Consolation is about?
    \item What is transnationalism?
\end{enumerate}

\subsubsection{Descriptive}

\begin{enumerate}
    \item Can you describe Ireland’s population loss process throughout the nineteenth century?
    \item What was the Young Ireland movement?
    \item What was the Land League?
    \item Can you describe the events that led to the Great Irish potato Famine?
    \item What were the main five consequences for Ireland of the Potato Famine of 1845?
    \item What was the Catholic Emancipation movement?
    \item Can you describe how was the Irish Act of Union passed in Parliament?
    \item Can you describe chain migration?
    \item What was the political state of Ireland in 1845?
    \item Who was Daniel O’Connell?
\end{enumerate}

\subsubsection{Integrative}

\begin{enumerate}
    \item What are the two most significant controversies that have developed in Irish diaspora studies in the past twenty-five years? Elaborate.
    \item What areas of study should Irish history researchers focus on in the future? Elaborate.
    \item What is the future of history as a discipline? Elaborate.
    \item Which geographic areas and historical time periods do you think should receive more or less attention based on what are the dominant themes in the current Irish Historiography? Elaborate.
    \item What elements do you consider to be crucial in teaching Irish diaspora studies, and what do you think needs to be revised? Elaborate.
    \item What is the future of migration studies? Elaborate. 
    \item Taking from Kerby Miller’s Emigrants and Exiles book, and for Donald Harman Akenson’s book The Irish in Ontario, what are the main arguments these historian make, how are these arguments similar and different, and what are contested subjects in their works? Elaborate.
    \item How does Irish migration to the United States differ from Irish migration to Canada? Elaborate.
    \item Were the Irish in the United States considered white? Elaborate.
    \item What were Irish migrants’ role in African slavery? Elaborate.
\end{enumerate}

\subsection{ChatGPT answer to Irish Potatoe Famine}
\label{app:chatgpt_answer_famine}

\answer{ChatGPT}: "The Great Irish Potato Famine, also known as the Great Hunger or the Irish Potato Famine, was a devastating period in Irish history that occurred between 1845 and 1852. It was a catastrophic event primarily caused by a combination of potato crop failure, economic policies, and socio-political factors. Here is a description of the events that led to the famine: Potato Dependence: The Irish population in the mid-19th century heavily relied on the potato as a staple food crop. The potato was affordable, easy to grow, and provided sufficient calories to sustain a large population. However, this heavy dependence on a single crop made the population vulnerable to any failures or diseases affecting the potato harvest. Potato Blight: In 1845, a strain of potato blight, caused by a fungal disease called Phytophthora infestans, spread across Ireland. The blight rapidly destroyed the potato crops, turning them into a black, rotting mass. This initial outbreak resulted in a significant reduction in the potato harvest, leading to food shortages and increasing fears of famine. British Economic Policies: Ireland was under British rule during this time, and British economic policies exacerbated the impact of the blight. British policies favored exporting Irish agricultural produce, such as grain and livestock, to Britain while leaving the Irish population heavily dependent on the potato as their main food source. The policies prioritized the interests of British landowners and exacerbated existing economic inequalities in Ireland. Limited Relief Efforts: The British government's response to the crisis was initially insufficient. Relief efforts were limited, and policies were implemented with a focus on maintaining economic stability rather than prioritizing the well-being of the Irish population. This lack of effective intervention worsened the situation and increased the suffering of those affected by the famine. Rapid Spread and Devastation: The potato blight continued to devastate subsequent potato crops in the years that followed. The impact was particularly severe in the west and southwest regions of Ireland, where the population was heavily reliant on potato cultivation. The loss of the potato harvest resulted in widespread hunger, malnutrition, and disease. Mass Emigration and Death: The combination of crop failure, lack of relief, and dire living conditions led to mass emigration and significant loss of life. Many Irish people, desperate to escape the famine's effects, left their homes and families to seek opportunities in other countries, particularly the United States and Canada. However, emigration was not an option for everyone, and the lack of adequate food and resources resulted in a devastating death toll, estimated to be around one million people, while millions more emigrated. The Great Irish Potato Famine had a profound and lasting impact on Ireland. It led to depopulation, cultural trauma, and a reevaluation of Irish-British relations. The famine exposed the socio-economic inequalities, poor relief systems, and inadequate governance during that time, and its effects continue to be remembered as a defining period in Irish history."

\section{Hallucination Example}
\label{app:hallucination_example}

\begin{quote}
    \question{Question 9 (Argumentative)}: "Can you summarize what David Fitzpatrick’s book Oceans of Consolation is about?" 
    
    \answer{GPT3}: "David Fitzpatrick's book "Oceans of Consolation" is a fundamental study and model for further editions of emigrant letters. It explores the personal letters sent to and from Irish migrants in Australia, providing full transcripts and examining each correspondent's local context. The book aims to understand the settlement of the Irish abroad in a comparative context and emphasizes the importance of faith in the lives of Irish migrant." 
    
    \answer{Falcon}: "Ocean of Consolation is a book about the history of the ocean and its impact on human culture. It explores the various ways the ocean has been perceived throughout history, from being seen as a source of wealth and power to being a place of spiritual significance. The book also examines the current state of the world's oceans and the threats they face due to human activity."
\end{quote}

\section{Irish Migration Studies Dataset}
\label{app:irish_migration_studies_dataset}

Akenson, Donald H. 1999. The Irish in Ontario : A Study in Rural History. 2nd ed. Montreal: McGill-Queen's University Press.

Barcia Paz, Manuel. 2008. Seeds of Insurrection : Domination and Resistance on Western Cuban Plantations 1808-1848. Baton Rouge: Louisiana State University Press.

Barclay, Katie. 2019. Men on Trial : Performing Emotion Embodiment and Identity in Ireland 1800-45. Manchester: Manchester University Press.

Bielenberg, Andy. 2013. The Irish Diaspora. Oxon England: Routledge.

Bravo Lozano, Cristina. 2018. Spain and the Irish Mission, 1609-1707. New York: Routledge.

Brehony, Margaret and Nuala Finnegan. 2019. Irlanda y Cuba: Historias Entretejidas = Irlanda \& Cuba : Entangled Histories. Cuba: Ediciones Bolona.

Brettell, Caroline. 2007. Constructing Borders. Lanham MD: Lexington Books.

Brettell, Caroline and James Frank Hollifield. 2023. Migration Theory: Talking Across Disciplines. New York NY: Routledge.

Brundage, David. 2016. Irish Nationalists in America: The Politics of Exile, 1798-1998. Oxford: Oxford University Press.

Bock, Gisela. 1989. “Women's History and Gender History: Aspects of an International Debate.” Gender \& History 7–30.

Bornstein, George. 2011. The Colors of Zion: Blacks, Jews, and Irish from 1845 to 1945. Cambridge: Harvard University
Press.

Bueltmann, Tanja; David T Gleeson; and Donald M MacRaild. 2012. Locating the English Diaspora 1500-2010. Liverpool: Liverpool University Press.

Butler, Judith. 1999. Gender Trouble: Feminism and the Subversion of Identity. New York: Routledge.

Butler, Judith. 2004. Undoing Gender. New York: Routledge.

Coldham, Peter Wilson. 1992. Emigrants in Chains : A Social History of Forced Emigration to the Americas of Felons Destitute Children Political and Religious Non-Conformists Vagabonds Beggars and Other Undesirables 1607-1776. Baltimore Md: Genealogical Pub.

Curry-Machado, Jonathan. 2011. Cuban Sugar Industry : Transnational Networks and Engineering Migrants in Mid-Nineteenth Century Cuba. 1st ed. New York: Palgrave Macmillan.

Curry-Machado, Jonathan. 2013. Global Histories Imperial Commodities Local Interactions. Houndmills Basingstoke Hampshire: Palgrave Macmillan.

Desley, Deacon; Penny Russell; and Angela Woollacott. 2010. Transnational Lives : Biographies of Global Modernity 1700-Present. Basingstoke England: Palgrave Macmillan.

Diaz, Maria Elena. 2022. The Virgin the King and the Royal Slaves of El Cobre : Negotiating Freedom in Colonial Cuba 1670-1780. Stanford CA: Stanford University Press.

Din, Gilbert C. 2016. An Extraordinary Atlantic Life: Sebasti\'an Nicol\'as Calvo De La Puerta y O'Farrill, Marqu\'es De Casa-Calvo. Lafayette LA: University of Louisiana at Lafayette Press. 

Duffy, Jennifer Nugent. 2013. Who’s your Paddy? Racial Expectations and the Struggle for Irish American Identity. New York: NYU Press.

Dunn, Richard and Rebekah Higgitt. 2015. Navigational Enterprises in Europe and Its Empires 1730-1850. Houndmills Basingstoke Hampshire: Palgrave Macmillan.

Dunn, Richard S. 2014. A Tale of Two Plantations : Slave Life and Labor in Jamaica and Virginia. Cambridge Massachusetts: Harvard University Press.

Dunn, Richard S and Gary B Nash. 2000. Sugar and Slaves : The Rise of the Planter Class in the English West Indies 1624-1713. Chapel Hill North Carolina: Published for the Omohundro Institute of Early American History and Culture Williamsburg Virginia by the University of North Carolina Press.

Evans, Nicholas and Angela McCarthy. 2022. Death in the Diaspora: British and Irish Gravestones. Edinburgh: Edinburgh University Press.

Fanning, Tim. 2018. Paisanos : The Irish and the Liberation of Latin America. Notre Dame Indiana: University of Notre Dame Press.

Fitzgerald Patrick, and B. K Lambkin. 2008. Migration in Irish History 1607-2007. Basingstoke England: Palgrave Macmillan.

Garcia Rodriguez Gloria. 2011. Voices of the Enslaved in Nineteenth-Century Cuba : A Documentary History. Chapel Hill: University of North Carolina Press.

Garner, Steven. 2007. ATLANTIC CROSSING, Atlantic Studies, 4:1, 117-132, DOI: 10.1080/14788810601179485. 

Giemza, Bryan Albin. 2013. Rethinking the Irish in the American South: Beyond Rounders and Reelers. University Press of Mississippi.

Gleeson, David T. 2013. The Green and the Gray : The Irish in the Confederate States of America. Chapel Hill: University of North Carolina Press.

Gleeson, David T. 2001. The Irish in the South 1815-1877. Chapel Hill: University of North Carolina Press.

Gleeson, David T. and Simon Lewis. 2014. The Civil War As Global Conflict : Transnational Meanings of the American Civil War. Columbia South Carolina: University of South Carolina Press.

Gray, Breda. 2004. Women and the Irish Diaspora. London: Routledge.

Gordon, Linda. 2022. The Great Arizona Orphan Abduction. Cambridge MA: Harvard University Press.

Guglielmo, Jennifer. 2010. Living the Revolution : Italian Women's Resistance and Radicalism in New York City 1880-1945. Chapel Hill: University of North Carolina Press. 

Hall, Catherine and Sonya O Rose. 2006. At Home with the Empire : Metropolitan Culture and the Imperial World. Cambridge UK: Cambridge University Press.

Hodgson, Kate. 2018. “Franco-Irish Saint-Domingue Family Networks Trans-Colonial Diasporas.” Caribbean Quarterly 434–51.

Horning, Audrey. 2013. Ireland in the Virginian Sea: Colonialism in the British Atlantic. Chapel Hill: University of North Carolina Press.

Houston, Cecil J. and William J. Smyth. 1990. Irish emigration and Canadian settlement: patterns, links, and letters. Toronto: U. of Toronto Press.

Jordan, Don, and Michael Walsh. 2008. White Cargo: The Forgotten History of Britain’s White Slaves in America.
New York: NYU Press.

Josiah, Barbara P. 2011. Migration Mining and the African Diaspora : Guyana in the Nineteenth and Twentieth Centuries (version 1st ed). 1st ed. New York: Palgrave Macmillan.

Kenny, Kevin. 2013. Diaspora : A Very Short Introduction. Oxford: Oxford University Press.

Kenny, Kevin. 1998. Making Sense of the Molly Maguires. New York NY: Oxford University Press.

Knight, Franklin W. 1970. Slave Society in Cuba during the Nineteenth Century. Madison: University of Wisconsin Press.

Kuethe, Allan J. 1986. Cuba 1753-1815: Crown Military and Society. 1st ed. Knoxville: University of Tennessee Press.

Ignatiev, Noel. 2012. How the Irish Became White. Hoboken: Taylor and Francis.

Ireland, John de Courcy. 1986. Ireland and the Irish in Maritime History. Dublin Ireland: Glendale Press.

Lorimer, Joyce. 2018. English and Irish Settlement on the River Amazon 1550-1646. Farnham: Routledge.

Mar, Lisa Rose. 2010. Brokering Belonging : Chinese in Canada's Exclusion Era 1885-1945. New York: Oxford University Press.

Marshall, Oliver. 2000. English-Speaking Communities in Latin America. New York: St. Martin's Press.

McCarthy, Angela. 2005. Irish Migrants in New Zealand 1840-1937 : 'the Desired Haven'. Woodbridge Suffolk U.K: Boydell Press.

McCarthy, Angela and Catharine Coleborne. 2012. Migration Ethnicity and Mental Health : International Perspectives 1840-2010. New York: Routledge.

McCarthy, Angela. 2011. Scottishness and Irishness in New Zealand Since 1840. Manchester: Manchester University Press.

McGaughey, Jane. 2020. Violent Loyalties: Manliness, Migration, and the Irish in the Canadas, 1798-1841. Liverpool University Press.

McMahon, Eileen M. 1995. What parish are you from? A Chicago Irish community and race relations. Lexington:
University Press of Kentucky.

McMahon, Timothy G., Michael de Nie, Paul Townsend. 2017. Ireland in an Imperial World. Citizenship, Opportunism, and
Subversion. Cambridge University Press.

Middleton, Stephen; David R. Roediger; and Donald M Shaffer. 2017. The Construction of Whiteness : An Interdisciplinary Analysis of Race Formation and the Meaning of a White Identity. Jackson: University Press of Mississippi.

Miller, Kerby A. 1985. Emigrants and Exiles: Ireland and the Irish Exodus to North America. New York: Oxford University Press.

Mohanram, Radhika. 2007. Imperial White : Race Diaspora and the British Empire. Minneapolis: University of Minnesota Press.

Morgan, Gwenda and Peter Rushton. 2013. Banishment in the Early Atlantic World: Convicts Rebels and Slaves. London: Bloomsbury.

Morton, Graeme, and David A. Wilson, eds. 2013. Irish and Scottish Encounters with Indigenous Peoples: Canada, the United States, New Zealand, and Australia. McGill-Queens University Press.

Mosse, George L. 1996. The Image of Man: The Creation of Modern Masculinity. New York \& Oxford: Oxford University
Press.

Murray, Edmundo.2005. Becoming Irlandés: Private Narratives of the Irish Emigration to Argentina, 1844-1912. Buenos
Aires.

Nash, Catherine. 2015. Genetic Geographies: The Trouble with Ancestry. Minneapolis: University of Minnesota Press.

Nelson, Bruce. 2012. Irish Nationalists and the Making of the Irish Race. Princeton: Princeton University Press.

O'Brien, Sarah. 2017. Linguistic Diasporas Narrative and Performance : The Irish in Argentina. Cham Switzerland: Palgrave Macmillan.

O'Callaghan, Sean. 2000. To Hell or Barbados : The Ethnic Cleansing of Ireland. Dingle: Brandon.

O'Connor, Thomas; Mary Ann Lyons. 2003. Irish Migrants in Europe After Kinsale 1602-1820. Dublin: Four Courts Press.

O'Kane, Finola and Ciaran O'Neill. 2023. Ireland Slavery and the Caribbean: Interdisciplinary Perspectives. Manchester: Manchester University Press.

Onkey, Lauren. 2010. Blackness and Transatlantic Irish Identity: Celtic Soul Brothers. New York: Routledge.

O'Neill, Peter D. 2017. Famine Irish and the American Racial State. New York: Routledge.

Prados-Torreiram Teresa. 2005. Mambisas : Rebel Women in Nineteenth-Century Cuba. Gainesville Fla: University Press of Florida.

Putnam Lara. 2013. Radical Moves : Caribbean Migrants and the Politics of Race in the Jazz Age. Chapel Hill: University of North Carolina Press.

Recio Morales, Oscar. 2010. Ireland and the Spanish Empire 1600-1825. Dublin: Four Courts Press.

Roberts, Daniel Sanjiv, Jonathan Jeffrey Wright. 2019. Ireland’s Imperial Connections, 1775-1947. Palgrave MacMillan.

Rose, Sonya O. 1992. Limited Livelihoods : Gender and Class in Nineteenth-Century England. London: Routhledge.

Rose, Sonya O. 2010. What Is Gender History? Cambridge UK: Polity.

Scott, Joan Wallach. 1996. Feminism and History. Oxford: Oxford University Press.

Scott, Joan Wallach. 2018. Gender and the Politics of History (version 30th anniversary edition) 30Th anniversary ed. New York: Columbia University Press.

Scott, Joan Wallach. 2019. Knowledge Power and Academic Freedom. New York: Columbia University Press.

Scott, Joan Wallach. 2012. The Fantasy of Feminist History. Durham: Duke University Press.

Smith, Raymond T. 1988. Kinship and Class in the West Indies : A Genealogical Study of Jamaica and Guyana. Cambridge: Cambridge University Press.

Stradling, R. A. 1994. The Spanish Monarchy and Irish Mercenaries : The Wild Geese in Spain 1618-68. Blackrock Co. Dublin: Irish Academic Press.

Valente, Joseph. 2011. The Myth of Manliness in Irish National Culture 1880-1922. Urbana Ill: University of Illinois Press.

\end{document}